\documentclass[twocolumn, prb, showpacs,preprintnumbers,amsmath,amssymb]{revtex4}

\usepackage{graphicx}
\usepackage{bm}

\usepackage{dcolumn}
\newcommand{\caf}{{CaFe$_{2}$As$_{2}$}}

\begin{document}

\title{CaFe$_{2}$As$_{2}$: a springboard to investigating Fe-pnictide superconductivity}

\author{D. A. Tompsett}
 \email{dat36@cam.ac.uk}
\author{G. G. Lonzarich}%
\affiliation{Cavendish Laboratory, University of Cambridge, Madingley Road, Cambridge CB3 0HE, UK}%

\date{\today}

\begin{abstract}
We present detailed electronic structure calculations for CaFe$_{2}$As$_{2}$.
We investigate in particular the `collapsed' tetragonal and orthorhombic regions of the
temperature-pressure phase diagram and find properties that distinguish CaFe$_{2}$As$_{2}$ from other
Fe-pnictide compounds. In contrast to the tetragonal phase of other Fe-pnictides the electronic
structure in the `collapsed' tetragonal phase of CaFe$_{2}$As$_{2}$ is found to be strongly
3D. We discuss the influence of these properties on the formation of superconductivity and in particular we find
evidence that both magnetic and lattice interactions may be important to the formation of superconductivity. We also find that the Local Spin Density Approximation is able to accurately predict the ordering moment in the low
temperature orthorhombic phase.
\end{abstract}

\pacs{74.70.Dd, 71.15.Mb, 74.25Jb}
\keywords{CaFe2As2, Fermiology, superconductor, antiferromagnet}
\maketitle

\section{\label{sec:H1}Introduction\protect}
The recent discovery of superconductivity in the doped iron pnictide
compounds and subsequent improvement in $T_{c}$ has generated significant
interest in uncovering the mechanisms responsible for this novel superconductivity.
In the 1,2,2 class of Fe-pnictide compounds including (Ca,Ba,Sr,Eu)Fe$_{2}$As$_{2}$,
superconductivity has been shown under pressure tuning
 for (Ba,Sr,Eu)Fe$_{2}$As$_{2}$\cite{Alireza1, Miclea1} with transition temperatures as high as 32K.
There have been several reports of the presence of superconductivity in the phase diagram of
\caf~ as it is pressure tuned from the orthorhombic phase into the collapsed tetragonal phase at low
temperatures\cite{Torikachvili1,Park1}. The determination of the presence of superconductivity has
recently been shown to be more complex by measurements
made using a Helium pressure medium\cite{Yu1}. It has also been suggested that superconductivity
results from the presence of a mixed phase intermediate between the collapsed tetragonal and
orthorhombic phases\cite{Lee1}.
Therefore, CaFe$_{2}$As$_{2}$ may be a material that is near the border for
the formation of superconductivity and as a result its electronic and magnetic structure
have the potential to provide critical information regarding the superconductivity of
these compounds.

\begin{figure}
\includegraphics[width=0.4\textwidth]{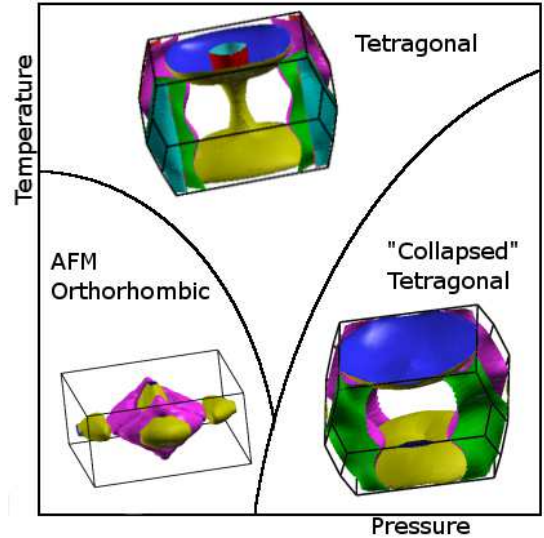}
\caption{\label{fig:PhaseWithFS} (Color online) Phase diagram of CaFe$_{2}$As$_{2}$ showing the Local Spin Density Approximation
 Fermi surface calculated in this study for each region. The dramatic changes in the Fermi surface as we introduce new
order by structural or magnetic phase changes illustrates coupling to the
electronic degrees of freedom. This is indicative of the need to consider both
magnetic and lattice interactions with the crystal structure
when discussing superconductivity in this class of compounds.}
\end{figure}

The Temperature-Pressure phase diagram for \caf~is unusual amongst Fe-pnictide compounds and is
shown schematically in Fig.~\ref{fig:PhaseWithFS}.
The presence of the `collapsed' tetragonal phase at low temperatures with the application of modest pressures
distinguishes this compound from other members of the 1,2,2 family\cite{McQueeney1}.
The low temperature orthorhombic antiferromagnetic phase that occurs
at low pressures is common to the 1,2,2 class and is thought to result from the formation
of spin density wave (SDW) itinerant magnetism\cite{Sebastian1}.
At high temperatures \caf~exists in a tetragonal structure and is non-magnetic\cite{Kreyssig1}.

In the calculations reported here we seek to consider the electronic structure of \caf throughout
the three key parts of the phase diagram. We investigate the mechanisms for the
novel phase transitions and superconductivity of this compound.
We also present predictions for
future quantum oscillation studies that may experimentally verify the electronic structure.

\section{\label{sec:H2}STRUCTURE AND METHOD\protect}
The electronic structures were determined by the Full-Potential LAPW method implemented
in WIEN2K\cite{Blaha1}. In all cases $R_{MT}^{min}K_{max} = 8$ was used. We have used the
Local Spin Density Approximation (LSDA) for the correlation functional except where
otherwise stated. We have used the experimental lattice parameters reported for each part of
the phase diagram. Further details will be discussed as is relevant to each calculation. Fermi surfaces were visualized with XCrysDen\cite{XCrysDen}.

In the following sections we consider each part of the phase diagram in turn. First, we
present the results for the ambient pressure tetragonal phase. These results place~\caf ~in a background
from which it may be referenced to other Fe-pnictide systems. We then consider the
collapsed tetragonal phase at $P$=0.63 GPa and contrast this electronic structure with the high temperature
tetragonal phase. Finally, we consider the antiferromagnetic orthorhombic state and investigate the prediction of the magnetic moment in this phase. With each of these
elements of the phase diagram in place, we are then in a position to qualitatively discuss their interplay in determining the
magnetic order and possible mechanisms for superconductivity in the system.

\section{\label{sec:HTetragonal}THE HIGH TEMPERATURE TETRAGONAL PHASE\protect}
\begin{figure}
\includegraphics[width=0.4\textwidth]{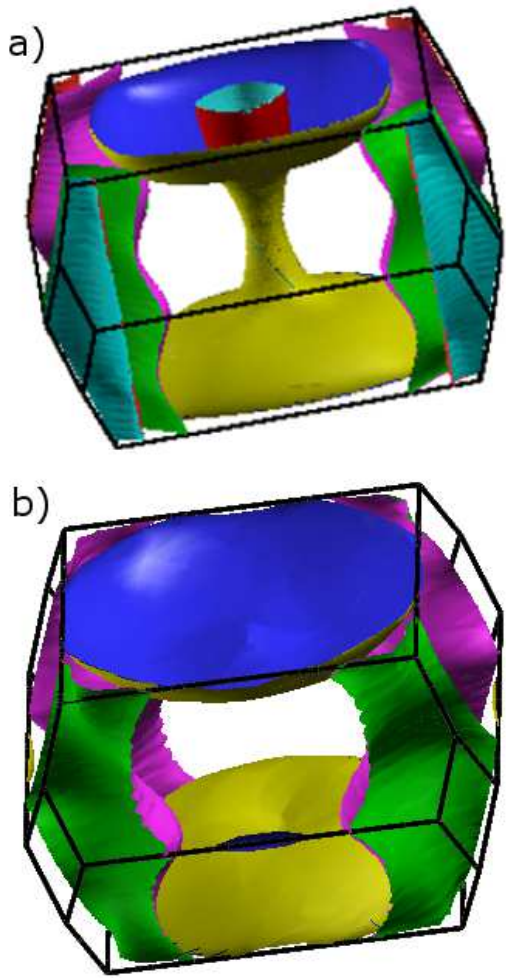}
\caption{\label{fig:CaFe2As2FSAmbCol} (Color online) a) The Fermi surface of CaFe$_{2}$As$_{2}$ in the high temperature tetragonal phase.
b) The Fermi surface of CaFe$_{2}$As$_{2}$ in the `collapsed' tetragonal phase at $P$=0.63 GPa\cite{Kreyssig1}. In contrast
to the Fermi surface in the ambient pressure tetragonal phase we find just two Fermi surface sheets.
One strongly 3D hole sheet that we refer to as the pillow as well as a strongly corrugated electron
cylinder at the zone corners. The 3D nature of the hole sheet makes the possibility of SDW instabilities
due to nesting in the Fermi surface less clear.}
\end{figure}

The crystal structure of the high temperature tetragonal phase of \caf~ is in the $I4/mmm$ tetragonal space group with lattice parameters
$a = 3.912$\AA ~and $b=11.667$\AA\cite{Nandi1}.
Calculations of the electronic structure of \caf~ in this phase have previously been presented for
comparison with X-ray photoelectron spectra\cite{Kurmaev1}. Here, we
present details, in particular the form of the Fermi surface, that are
important as a point of comparison between \caf~ and other Fe-pnictide systems as well
as for comparison with \caf~ in its collapsed tetragonal structure. We have calculated the
electronic structure using 31$\times$31$\times$31 k-points in the full Brillouin zone and the resulting
Fermi surface is shown in Fig.~\ref{fig:CaFe2As2FSAmbCol}(a). During the calculation we relaxed the lattice
position of the As atom, $z_{As}$, from its experimental position of $z_{As}=0.3665$ to $z_{As}=0.353(0)$ under the LSDA. The Fermi surface obtained is similar to that found in previous calculations for (Ba,Sr)Fe$_{2}$As$_{2}$\cite{Singh1,Sebastian1}.

It is intriguing to note the similarity between this Fermi surface and that of MgB$_{2}$ which also possesses
two concentric warped cylinders at the zone corners as well as a flat 3D pocket in a similar position to the flared
section of the dumbell surface. The primary difference is the lack of a further 3D pocket about $\Gamma$ which occurs
in MgB$_{2}$, but not in \caf. In MgB$_{2}$ it is the two outer cylinders which couple to the lattice to form
superconductivity\cite{Mazin2}.

\section{\label{sec:CTetragonal}THE `COLLAPSED' TETRAGONAL PHASE\protect}
In the collapsed tetragonal phase the crystal maintains the $I4/mmm$ symmetry of the high temperature tetragonal
phase, but undergoes a dramatic reduction in the c-axis lattice parameter of approximately 6\% while
simultaneously the in-plane dimension increases by approximately 2\%. The unit cell is effectively squashed.
At $P$=0.63 GPa the lattice parameters are
$a = 3.9780(1)$\AA ~and $b=10.6073(7)$\AA\cite{Kreyssig1}.
We have calculated the electronic structure for this phase using a 39$\times$39$\times$39 k-point grid.
The resulting Fermi surface is shown in Fig.~\ref{fig:CaFe2As2FSAmbCol}(b). In Table~\ref{tab:tableColFreqs} we also present the quantum oscillation frequencies and associated band masses for \textbf{B} parallel
to c-axis. Experimental comparison with these values has the ability to verify the electronic structure
obtained in our calculations and the experimental mass enhancement can indicate the strength of correlations in
the system.
\begin{table}
\caption{\label{tab:tableColFreqs} Theoretical quantum oscillation frequencies and band masses
for the collapsed tetragonal phase with the magnetic field, \textbf{B}, parallel
to the c-axis.}
\begin{ruledtabular}
\begin{tabular}{ccc}
Orbit\footnote{$\alpha$ denotes the warped cylinder around the zone corner in Fig.~\ref{fig:CaFe2As2FSAmbCol}(b), $\beta$ is the large pillow surface.}&Frequency (kT) & Band Mass ($m_{e}$)\\
\hline
$\alpha_{1}$ & 3.64 & 1.25\\
$\alpha_{2}$ & 6.02 & 2.17\\
$\alpha_{3}$ & 7.19 & 2.15\\
$\beta$ & 16.0 & 2.40\\
\end{tabular}
\end{ruledtabular}
\end{table}

A critical observation of this investigation is the dramatic difference between
the Fermi surfaces of the collapsed tetragonal phase and the high temperature tetragonal phase. The difference involves two
key alterations. Firstly, in the collapsed phase we find a Fermi surface that is composed
of just two sheets. Secondly, due to the further reduction in the c-axis lattice parameter the hole pocket
is now very 3D. Therefore,
the electronic structure of this collapsed tetragonal phase is distinct from both the high temperature tetragonal
phase and that of other members of the Fe-pnictide family.

It has been suggested that magnetically mediated Cooper pairing is more robust in 2D systems\cite{Monthoux1} and therefore the 3D Fermi
surface of the collapsed tetragonal phase may give rise to lower superconducting transition temperatures than similar 2D systems. In
fact if we consider the pressure induced superconducting transitions of the (Ba,Sr,Eu)Fe$_{2}$As$_{2}$ family then we find support
for this idea. (Ba,Sr,Eu)Fe$_{2}$As$_{2}$ show maximal pressure induced transition temperatures of approximately 32K, 28K and 29K respectively\cite{Alireza1, Miclea1}. Each of these compounds shows a more 2D paramagnetic electronic structure around the superconducting transition than
we have found here for \caf. Therefore, the explanation for the lower transition temperature of 12K\cite{Torikachvili1,Park1}, or absence of superconductivity
as reported by Lee \textit{et al.}\cite{Lee1}, may rely on the different dimensionality of the Fermi surface in this collapsed
tetragonal region.

The dramatic change in Fermi surface topology in moving through the
transition from the high temperature to the collapsed tetragonal phase is produced by a reduction in lattice
volume of just 5\%. This is indicative of a significant coupling between the electronic and lattice degrees of
freedom. A phononic influence in combination with magnetic interactions may therefore be active in
producing superconductivity in these compounds as has been suggested for BaNi$_{2}$As$_{2}$\cite{Singh4}.

\section{\label{sec:Orthorhombic}THE ORTHORHOMBIC PHASE\protect}

In the low temperature and pressure region of its phase diagram ~\caf~ is
antiferromagnetic (AFM) in the $Fmmm$ orthorhombic structure. The lattice parameters are
a=5.506(2)\AA, b=5.450(2)\AA, and c=11.664(6)\AA ~and $z_{As}=0.36642(5)$\cite{Goldman1}. The magnetic moment is 0.80(5)$\mu_{B}$/Fe\cite{Goldman1}.
The conventional magnetic unit cell is the same as that of the crystal structure, but the symmetry
is reduced due to the magnetic order. 

We have evaluated the electronic and magnetic structure for the experimentally determined magnetic
order\cite{Goldman1} using both the Generalized Gradient Approximation of Perdew-Burke-Ernzerhof (GGA) and the LSDA correlation functionals. For all calculations in this
antiferromagnetic phase we have used 23$\times$23$\times$10 k-points in the Brillouin zone.

First, we applied the GGA to the AFM structure. Relaxing the As position within the AFM calculation results in, $z_{As}=0.3659$, which matches the experimental value closely, but the predicted
moment of 1.84$\mu_{B}$/Fe overestimates the experimental value by a factor of more than 2. Overestimates of this scale for the
magnetic moment in Fe-pnictides while applying the GGA have been widely reported\cite{Singh2}.

\begin{figure}[h]
          \includegraphics[width=.5\textwidth]{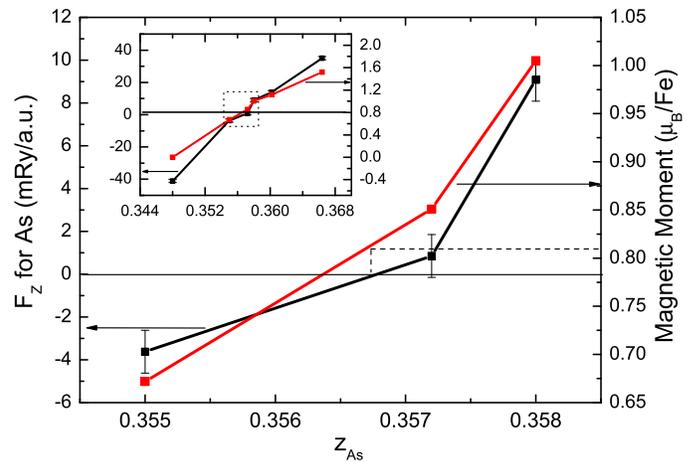}
     \caption{(Color online) Plot of the force on the As atom(black) and the magnetic moment of Fe(red) versus the lattice parameter $z_{As}$
     in~\caf. The forces and magnetic moments were determined under the Local Spin Density Approximation.  The main figure shows the region about $F_{Z}=0$
     where the As position is relaxed into its equilibrium structure. The inset is a
     view of this plot over a wider scale of $z_{As}$. The solid horizontal line indicates
     where the force $F_{Z}=0$. The dashed lines in the main figure are to illustrate the value of the moment
     in $\mu_{B}$/Fe that occurs when $F_{Z}=0$.}
     \label{fig:ForceMomAll}
\end{figure}

Secondly, we applied the LSDA to the antiferromagnetic structure while relaxing the value $z_{As}$. We find that
the relaxation of the As coordinate is significant, from $z_{As}=0.3664$ experimentally to $z_{As}=0.3567$. In Fig.~\ref{fig:ForceMomAll} we have plotted both the force on the As atom, $F_{z}$, and the magnetic moment of Fe
 as a function of its relaxation to the plane of the Fe atoms. Very importantly, this figure shows that the moment
 approaches $0.81\pm0.02\mu_{B}$/Fe as $F_{z}$ tends towards zero ie. when the As is in its
 crystallographic equilibrium position. This is in close agreement with the experimental reported moment of 0.80(5)$\mu_{B}$/Fe\cite{Goldman1}. To confirm this result we have applied the same method to SrFe$_{2}$As$_{2}$
 and BaFe$_{2}$As$_{2}$. The results are summarized in Table~\ref{tab:tableOrtMoms} and give strong agreement with experiment.

\begin{table}
\caption{\label{tab:tableOrtMoms} Comparison between the magnetic moment of 1,2,2 compounds predicted by the LSDA
when the As position is relaxed in the antiferromagnetic phase and the experimental values\cite{Goldman1, Zhao2, Tegel1, Huang1}.}
\begin{ruledtabular}
\begin{tabular}{ccccc}
 & \multicolumn{2}{c}{LSDA} & \multicolumn{2}{c}{Experimental}\\
Compound & $\mu$ ($\mu_{B})/Fe\footnote{The errors given are calculated from the convergence of the
force on the As atom.}$ & $z_{As}$ & $\mu$ ($\mu_{B})/Fe$ & $z_{As}$\\
\hline
\caf &               0.81$\pm0.02$ & 0.3567 & 0.80(5) & 0.36642(5) \\
SrFe$_{2}$As$_{2}$ & 0.97$\pm0.03$ & 0.3507 & 0.94(4)\footnote{More recent measurements\cite{Kaneko1} have found $\mu=1.01\mu_{B}/Fe$.} & 0.3612(3)\\
BaFe$_{2}$As$_{2}$ & 0.86$\pm0.02$ & 0.3444 & 0.87(3) & 0.35406(7)\\
\end{tabular}
\end{ruledtabular}
\end{table}

\begin{figure}
\includegraphics[width=0.4\textwidth]{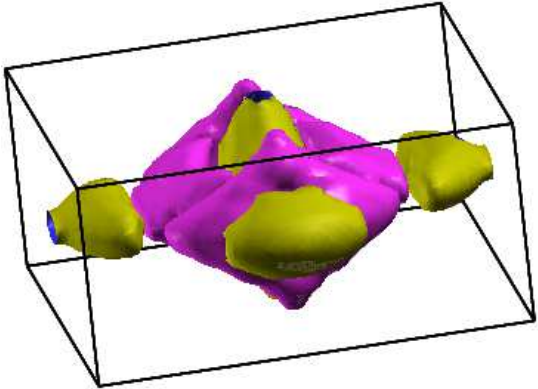}
\caption{\label{fig:AFMGeoOptLDAFS} (Color online) The Fermi surface of CaFe$_{2}$As$_{2}$ calculated
using the Local Spin Density Approximation with the As coordinate relaxed in the antiferromagnetic phase.}
\end{figure}
It has been suggested that
overestimations of the moment by Density Functional Theory may indicate quantum spin fluctuations. Such
effects are not fully accounted for under Density Functional Theory which is mean-field in nature. Our study, however, suggests
that a relaxation of the internal coordinates to the LSDA equilibrium position for As in the AFM phase is sufficient to permit
the accurate prediction of the magnetic moment. In some previous work the As coordinate has been relaxed
in a non-magnetic calculation\cite{Kreyssig1, Mazin1} which has resulted in overestistimates of the moment of approximately 1.5$\mu_{B}$/Fe. Other work has shown the presence of a smaller moment under the LSDA in LaFeAsO\cite{Yildirim1} and BaFe$_{2}$As$_{2}$\cite{Singh1}, but with less accurate correspondence with experiment than is shown in this study.
Together, these findings suggest that the preceding overestimation of the moment should not be used as evidence for zero-point fluctuations in this strongly antiferromagnetic material.

The Fermi surface of ~\caf~ for the LSDA calculation is shown in Fig.~\ref{fig:AFMGeoOptLDAFS}. Most of the Fermi surface is gapped away by the additional ordering vector to leave the small pockets shown. Analysis of the associated quantum oscillation frequencies shows broad
agreement with the recent experimental results of Harrison \textit{et al.}\cite{Harrison1}

The results presented here have the potential to motivate further studies of the
magnetic structure in these compounds that are both qualitatively and quantitatively
accurate.

\section{\label{sec:SummaryConclusions}SUMMARY AND CONCLUSIONS\protect}
The phase diagram of Fig.~\ref{fig:PhaseWithFS} illustrates that the \caf~Fermi surface changes
dramatically through magnetic and structural phase transitions.
This is suggestive that both magnetic and lattice interactions should be considered in
the formulation of any model of superconductivity in this compound.

The greater three dimensionality of the
collapsed phase Fermi surface suggest that pairing due to magnetic interactions may be weaker and this
may explain the lower superconducting $T_{c}$\cite{Torikachvili1, Park1}, or absence of superconductivity\cite{Yu1}, in ~\caf ~when compared to
(Ba,Sr,Eu)Fe$_{2}$As$_{2}$ under pressure.

Predictions for quantum oscillation experiments on the collapsed tetragonal phase have been given.
We have also demonstrated that the magnetic moment in the AFM orthorhombic phase may be accurately
calculated under the LSDA if the As coordinate is relaxed within an AFM calculation.

\bibliography{cafv5}

\end{document}